\documentclass[12pt,preprint]{aastex}

\def\gsim {\ifmmode {\buildrel>\over\sim}               
 \else {\lower.6ex\hbox{$\buildrel>\over\sim$}}\fi}
\def\lsim {\ifmmode {\buildrel<\over\sim}               
 \else {\lower.6ex\hbox{$\buildrel<\over\sim$}}\fi}

\slugcomment{Not to appear in Nonlearned J., 45.}

\shorttitle{Detection of CO$^+$ in M~82}
\shortauthors{Fuente et al.}

\begin{document}


\title{Detection of CO$^+$ in the nucleus of M82}


\author{A. Fuente$^1$, S. Garc\'{\i}a-Burillo$^1$, M. Gerin$^2$, J.R. Rizzo$^{3,1}$, 
A. Usero$^{1}$,D. Teyssier$^4$, E. Roueff$^5$, J. Le Bourlot$^5$}


\altaffiltext{1}{Observatorio Astron\'omico Nacional (OAN), Apdo. 112,
E-28803 Alcal\'a de Henares (Madrid), Spain}
\altaffiltext{2}{Laboratoire d'Etude du Rayonnement et de la Matiere en la Astrophysique, UMR 8112, CNRS, Ecole
Normale Superieure et Observatoire de Paris, 24 rue Lhomond, 75231 Paris Cedex 05, France}
\altaffiltext{3}{Departamento de F\'{\i}sica, UEM, Urb. El Bosque, E-28670 Villaviciosa 
de Od\'on (Madrid), Spain}
\altaffiltext{4}{Herschel Science Centre, European Space Astronomy Centre (ESAC), Urb. Villafranca del Castillo
P.O. Box 50727, E-28080 Madrid, Spain}
\altaffiltext{5}{LUTH and UMR 8102 du CNRS, Observatoire de Paris, 92195 Meudon Cedex, France}


\begin{abstract}
We present the detection of the reactive ion CO$^+$ towards the prototypical starburst galaxy
M~82. This is the first secure detection of this short-lived ion in an external galaxy. Values of 
[CO$^+$]/[HCO$^+$]$>$0.04 are measured across the inner 650~pc of the nuclear disk
of M~82. Such high values of the
[CO$^+$]/[HCO$^+$] ratio had only been previously measured towards the atomic peak
in the reflection nebula NGC~7023. This detection corroborates
that the molecular gas reservoir in the M~82 disk is heavily affected by the 
UV radiation from the recently formed stars. Comparing the column densities measured in
M~82 with those found in prototypical Galactic photon-dominated regions (PDRs), 
we need $\sim$20 clouds along the 
line of sight to explain our observations. We have completed 
our model of the molecular gas chemistry in the M~82 nucleus.
Our PDR chemical model 
successfully explains the [CO$^+$]/[HCO$^+$] ratios measured
in the M~82 nucleus but fails by one order of magnitude to explain 
the large measured CO$^+$ column densities ($\sim$1--4$\times$10$^{13}$~cm$^{-2}$).
We explore possible routes to reconcile the chemical model and the observations.

\end{abstract}




\section{Introduction}
M~82 is one of  the nearest and brightest starburst galaxies. 
Located at a distance of 3.9~Mpc, and with a bolometric luminosity
of 3.7$\times$10$^{10}$~L$_{\odot}$, it has been 
extensively studied in many molecules. 
Several studies reveal that the starburst has heavily influenced the
interstellar medium in M~82 by producing high UV and cosmic ray fluxes. 
Recent interferometric observations at millimeter wavelengths \citep{gar01} 
show that while the chemistry of the molecular gas in the disk-halo interface 
is dominated by shocks, 
the chemistry of the molecular gas in the M~82 disk seems to
be dominated by the intense UV flux. \citet{gar02} obtained
a high-angular-resolution image showing widespread 
enhanced HCO abundance ([HCO]/[H$^{13}$CO$^+$]$\sim$3.6)
across the whole M~82 disk which was interpreted in terms of a giant
PDR of 650~pc size.  Fuente et al. 2005 (hereafter Paper I)
observed a selected set of PDR tracers 
(CN,C$_2$H,HOC$^+$ and c-C$_3$H$_2$) 
in three positions across the M~82 disk.
They measured [CN]/[HCN]$\sim$5 in the inner 650~pc galaxy disk. 
This large value of the [CN]/[HCN] ratio is only reached in 
the most heavily exposed layers of a PDR to the UV radiation
\citep{fue93,fue97,ste95,bog05}.
Furthermore, we detected the HOC$^+$ 1$\rightarrow$0 line and obtained 
{\bf an} [HCO$^+$]/[HOC$^+$] ratio of $\sim$ 40. 
Such a low [HCO$^+$]/[HOC$^+$] ratio
is only expected in highly ionized molecular gas (X(e$^-$)$>$10$^{-5}$) either by 
UV photons (PDRs) and/or X-rays (XDRs) \citep{fue03,riz03,use04}.

In this paper we report the CO$^+$ detection in the nucleus of M~82.
This is the first secure CO$^+$ detection in an extragalactic object since previously
it has only been tentatively detected towards the active galactic nucleus (AGN)
Cygnus A \citep{fue00}. 
CO$^+$ is a reactive ion that can only survive
in the highly ionized layers of photon-dominated and
X-rays dominated regions (PDRs and XDRs) (see e.g. Sternberg \& Dalgarno 1995). 
In our Galaxy, CO$^+$ has only been detected in a handful of objects which are
well known prototypical PDRs 
(NGC 7027 and M17 SW: Latter et al. 1993,
Fuente et al. 2003; the Orion Bar: St{\"o}rzer et al. 1995, Fuente et al. 2003; 
NGC 7023:  Fuente \& Mart\'{\i}n-Pintado 1997, 
Fuente et al. 2003; Mon R2 and G29.96-0.02: Rizzo et al. 2003;
S140 and NGC~2023: Savage \& Ziurys 2004). 
In fact, CO$^+$ may be the best molecular tracer of the outermost layers (A$_v$$>$2~mag) 
of PDRs. In contrast with other molecular PDR tracers like CN and HOC$^+$,
CO$^+$ is exclusively formed in these layers by reactions of C$^+$
with OH. The formation of the chemically related ion HOC$^+$ is favored
in this region but can also be formed at a smaller rate in molecular clouds.
In addition to the CO$^+$ detection, we present observations of the 
high excitation HCO$^+$ 3$\rightarrow$2, HOC$^+$ 3$\rightarrow$2 
and the CH$_3$OH 5$_{k,k'}$$\rightarrow$4$_{k,k'}$ lines.

\section{Observations and Results}
The observations were carried out in June and November 2004 with the IRAM 30m 
radiotelescope at Pico de Veleta (Spain). We used 2 SIS receivers tuned in 
single-sideband mode in the 1 mm band. 
The observed transitions are: HCO$^+$ 3$\rightarrow$2,
HOC$^+$ 3$\rightarrow$2, CO$^+$ N=2$\rightarrow$1 J=5/2$\rightarrow$3/2 and
J=3/2$\rightarrow$1/2, and CH$_3$OH 5$_2$$\rightarrow$4$_2$ and
5$_{-2}$$\rightarrow$4$_{-2}$ E . In Fig. 1 we present the observed
spectra and the Gaussian fits are shown in Table 1.  
The intensity scale is main beam brightness temperature. 
The forward efficiency ($\eta_{ff}$), main-beam efficiency ($\eta_{MB}$)
and half-power beam width (HPBW) of the telescope 
are 0.91,0.52 and 11$''$ at 235 GHz and 0.88, 0.46 and 9$''$ at 260 GHz.
Pointing was checked every two hours by observing a reference source.
We observed three positions across the M~82 disk in the CO$^+$ lines: 
the nucleus (RA(2000): 09$^h$55$^m$51.9$^s$,  
Dec(2000): 69$^{o}$04$'$47.11$''$) (hereafter
referred to as {\it Center}) and the two peaks 
in the HCO emission [offsets ($+$14$''$,$+$5$''$) 
and ($-$14$''$,$-$5$''$) hereafter referred to as 
{\it E} and {\it W} respectively]. 
Only {\it E} and {\it Center} were observed in
the HCO$^+$ 3$\rightarrow$2 and HOC$^+$ 3$\rightarrow$2 lines. 
The HOC$^+$ 3$\rightarrow$2 line has been detected towards {\it E}. 
This detection constitutes a further corroboration of the 
HOC$^+$~1$\rightarrow$0 detection reported in Paper I.
The  CH$_3$OH line was only observed towards {\it Center}.

\subsection{CO$^+$ detection}

CO$^+$ has a $^2$$\Sigma$ ground electronic state in which each rotational level is split into
two fine-structure levels with J=N$\pm$1/2. The N=1$\rightarrow$0 rotational line is heavily obscured
by the O$_2$ line at 118 GHz and cannot be observed from ground-based telescopes. 
The most intense transitions of the N=2$\rightarrow$1 rotational spectrum are J=5/2$\rightarrow$3/2
at 236.062 GHz and J=3/2$\rightarrow$1/2 at 235.789 GHz. Since they are very close in
frequency, the two lines can be observed simultaneously. In the optically thin limit,
the intensity ratio I(236.062)/I(235.789) is 1.8. The detection of the two lines
with the expected intensity ratio provides supporting evidence of the reality of the  CO$^+$ detection.
 
The CO$^+$  N=2$\rightarrow$1 J=5/2$\rightarrow$3/2  line
has been detected  towards
{\it E} ($>$4$\sigma$) and {\it Center} ($>$8$\sigma$) and 
very tentatively ($\sim$3$\sigma$) towards {\it W}.  
Furthermore, we have a $\sim$3$\sigma$ detection of the weakest 
component towards {\it Center}. 
The 236.062 GHz CO$^+$ line is blended with the 5$_{-2,4}$$\rightarrow$4$_{-2,3}$  and
5$_{2,3}$$\rightarrow$4$_{2,2}$ E lines
of $^{13}$CH$_3$OH (at 236.062 and 236.063GHz respectively). To confirm the
CO$^+$ detection and estimate the possible contamination of the 13-methanol lines, we have
observed towards {\it Center} the same transitions of the abundant isotope $^{12}$CH$_3$OH
and obtained T$_{MB}$$\sim$10~mK. 
Assuming a $^{12}$C/$^{13}$C ratio $>$50 \citep{mao00}, the intensity of
the CH$_3$OH line should be $<$0.2~mK towards {\it Center}, i.e., at least
a factor of 30 lower 
than the intensity observed in the CO$^+$ spectrum. Therefore, we can conclude 
that the detected emission
at 236.062 GHz corresponds to the J=3/2$\rightarrow$1/2 line of CO$^+$. 

\subsection{Column density ratios}
The physical conditions of the molecular gas
are estimated by fitting the intensities of the 
H$^{13}$CO$^+$ 1$\rightarrow$0 and HCO$^+$ 3$\rightarrow$2
lines using an LVG code.
For these calculations, we assume T$_k$=50~K \citep{wei01}
and  [HCO$^+$]/[H$^{13}$CO$^+$]=89.
The $^{12}$C/$^{13}$C ratio in the M~82 nucleus is not
well known. \citet{mao00} derived
a $^{12}$C/$^{13}$C ratio between 50 and 75 from multiline
CO observations using an LVG code. 
However they obtained a larger value of the $^{12}$C/$^{13}$C ratio
when applying a PDR model to the same data. For consistency with Paper I, we adopt 
the canonical value [HCO$^+$]/[H$^{13}$CO$^+$]=89.
The derived molecular hydrogen densities 
are 3--8$\times$10$^4$~cm$^{-3}$.  
These densities are in agreement with those derived from the  
CN and HCN lines in Paper I within the uncertainties inherent to
this kind of calculations. Assuming these densities and 
an emission size of 6$"$ for  {\it E}, {\it W} and {\it Center}
we obtain the H$^{13}$CO$^+$ and HOC$^+$ column 
densities shown in Table 2. 
The size of the emission towards {\it E}, 
{\it W} and {\it Center} has been derived from 
the interferometric HCO and H$^{13}$CO$^+$
images published by \citet{gar02} (see Fig.~1).
To calculate the CO$^+$ column density, we assume
optically thin emission and use the 
Local Thermodynamic Equilibrium (LTE) approximation with
T$_{rot}$=10~K. This rotation temperature has been estimated
using a LVG code applied to a linear molecule with the same dipole
moment as CO$^+$ and n(H$_2$)=10$^5$~cm$^{-3}$.
Assuming a size of 6$"$ for the CO$^+$ emission we 
derive CO$^+$ column densities 
$\sim$1--4$\times$10$^{13}$~cm$^{-2}$
across the M~82 disk.

We can calculate the [HCO$^+$]/[HOC$^+$]
ratio towards {\it E} using the HCO$^+$ 3$\rightarrow$2 and 
HOC$^+$~3$\rightarrow$2 lines. Since both lines
lie at the same frequency and have similar dipole moments, 
[HCO$^+$]/[HOC$^+$]$\approx$I(HCO$^+$~3$\rightarrow$2)/I(HOC$^+$~3$\rightarrow$2) 
ratio. We obtain [HCO$^+$]/[HOC$^+$]$\approx$48  in excellent
agreement with our previous estimates (Paper I).

In Table~2 we compare the column
density ratios in M~82 with 
those measured in some prototypical Galactic PDRs.  
The [CO$^+$]/[HCO$^+$] ratio is larger than 0.04 all across the
M~82 disk. 
This is one of the largest values of the
[CO$^+$]/[HCO$^+$] ratio measured thus far.
Values of the [CO$^+$]/[HCO$^+$] ratio larger than
0.01 are only found towards the atomic peaks in 
prototypical Galactic
PDRs, and such a high value of the [CO$^+$]/[HCO$^+$] ratio
has only been measured in the
most exposed layers of the PDR associated with the reflection
nebula  NGC~7023.
This result is consistent with the values of the
[HCO$^+$]/[HOC$^+$] and [CN]/[HCN] ratios 
previously measured in the M~82
disk (Paper I). We estimated  [HCO$^+$]/[HOC$^+$]$\sim$40 and 
[CN]/[HCN]$>$5 across the M~82 nucleus. 
These values are also similar to those measured in the PDR peak
towards NGC~7023. 
The very favorable geometry of the PDR associated
with this reflection nebula allowed us to detect the  outermost layers
of the PDR \citep{fue93,fue96a}. Our observations suggest that the bulk of the
dense molecular gas in M~82 is surviving in a similar environment to that found
in the HI/H$_2$ transition layer of this photon-dominated region.

We can also compare the CO$^+$, HOC$^+$ and CN column densities 
in M~82 with those derived in Galactic PDRs, although the derived
column densities
are more uncertain than the column density ratios since the former
depend on the assumed beam filling factor.
In Galactic PDRs, the CO$^+$ column densities
are quite uniform taking a value of $\sim$10$^{12}$~cm$^{-2}$ in all star
forming regions regardless of the incident UV field in 
a range of 3 orders of magnitude. 
Furthermore, the [CO$^+$] /[HOC$^+$] ratio
is $\sim$0.5--9 \citep{riz03,sav04}. 
The [CO$^+$]/[HOC$^+$] ratio measured in M~82 is similar to
those found in the PDRs associated with star forming
regions in our Galaxy (see Table 2). This suggests a similar 
CO$^+$/HCO$^+$/HOC$^+$ chemistry and a similar
origin for the reactive ions. 
Since the CO$^+$  column density is a factor
20-40 larger in M~82 than in Galactic PDRs, we
need about 20-40 PDRs 
along the line of sight to account for our observations. 
In the scenario of clouds  immersed in a pervasive 
UV field, this implies about 10--20 clouds along the line of sight
which is a reasonable number for an edge-on galaxy. 

\section{Chemical Model}

To have a deeper insight
into the physics and chemistry of the molecular clouds in M~82,
we have modeled their chemistry using the PDR Le Bourlot's semi-infinite
plane parallel model (Le Bourlot et al. 1993) and the same physical 
conditions as in Paper I:  G$_0$=10$^4$ in units of Habing field, 
n=n$_H$+2$\times$n$_{H_2}$=4$\times$10$^5$~cm$^{-3}$
and a cosmic ray flux of 4$\times$10$^{-15}$~s$^{-1}$. 
As discussed in Paper I, this model
explains with reasonable success all the molecular 
abundances observed 
in M~82 and should therefore account for our CO$^+$ detection.  
 
The model results for CO$^ +$ and HCO$^+$ are shown in Fig.~2.
The N(CO$^+$)/N(HCO$^+$) ratio (hereafter $r_{CO^+}$) is very high 
($>$0.1) for A$_v$$<$0.5~mag. Then, $r_{CO^+}$ remains
constant and equal to $\sim$0.1 until A$_v$$<$5~mag.
This is because in this range of visual extinction, 0.5$<$A$_v$$<$4.5~mag,
$r_{CO^+}$ is determined by the CO$^+$ and HCO$^+$ abundances
in the most external layers (A$_v$$<$1~mag) of the PDR.
For higher 
extinctions, $r_{CO^+}$ decreases
because of the rapid increase of the HCO$^+$ abundance.
In our plane-parallel model,
the values of $r_{CO^+}$  
observed across the M~82 nucleus ($r_{CO^+}$$>$0.04) 
are found only for A$_v$$<$6.5~mag 
(see Fig.~2). In an external galaxy, one does not expect to have
a single PDR but a population of clouds (or cloudlets) immersed
in an intense UV field (Paper I, Garc\'{\i}a-Burillo et al. 2002).
In this scenario, our model results imply that the individual cloudlets 
have N$_{tot}$$\lsim$1.3~10$^{22}$~cm$^{-2}$.
Thus, our CO$^+$ observations corroborate the scenario
for the M~82 nucleus proposed in Paper I of a highly 
fragmented interstellar medium
in which the dense cores (n$\sim$4$\times$10$^5$~cm$^{-3}$, 
N$_{tot}$$\lsim$1.3$\times$10$^{22}$~cm$^{-2}$ ) 
are bathed by an intense UV field (G$_0$=10$^4$ Habing
fields). 
  
Thus far, we have only compared the observed and model
predicted molecular column density ratios. 
We can also compare the molecular column densities.
The predicted CO$^+$ column density is 
$\sim$3$\times$10$^{10}$~cm$^{-2}$
for A$_v$=6.5~mag. 
This value is  a factor of 20--40 lower
than the CO$^+$ column densities observed in the
prototypical Galactic PDRs. Furthermore it is 3 orders of magnitude
lower than the CO$^+$ column densities measured in
M~82 and an unrealistic large number of PDRs 
along the line of sight would be required
to explain our CO$^+$ observations. To have a deeper insight into
the cause of this discrepancy between theoretical
predictions and the observations, we have also compared 
the predicted CN column densities with the observed ones
(see Table 1). Like CO$^+$, 
CN is a good tracer of PDRs and
can be used to estimate the number of PDRs along
the line of sight. 
Our model predicts N(CN)$\sim$1.5$\times$10$^{14}$~cm$^{-2}$ for
A$_v$$\sim$6.5~mag. This value agrees within a factor of 2
with those observed in 
prototypical Galactic PDRs like the Orion Bar and NGC~7023.
Comparing with the CN column densities observed in the
M~82 nucleus, we need about 20-40 individual cloudlets 
along the line of sight to account for our
observations (see also Boger \& Sternberg 2005).
This is a reasonable number of cloudlets for an edge-on galaxy. 
Thus, there is a 
reasonable agreement between model
predictions and observations for CN in both, Galactic
and extra-galactic PDRs. However, 
the chemical model falls short by 
more than one order of magnitude of accounting for the
CO$^+$ column densities measured in Galactic PDRs and
the M~82 nucleus.
 
The failure of chemical models to account for the observed
reactive ions
column densities is a long standing problem \citep{bla98,fue00}.
The chemistry of reactive ions is very sensitive
to the gas physical conditions in the HI/H$_2$ transition layer. 
In particular, CO$^+$ is mainly produced via the reaction
C$^+$+OH$\rightarrow$CO$^+$+H. The production 
of OH is very dependent on the temperature as O+H$_2$ may 
come into play at the H/H$_2$ transition region when H$_2$ is abundant and 
the temperature is still a few hundred K. The corresponding 
endothermicity is about 3000~K.
An increase in the gas kinetic temperature in the HI/H$_2$ interface could have 
a dramatic effect in the CO$^+$ production. 

There are several observational evidences
that suggest that chemical models fail to predict the gas kinetic temperature in the
HI/H$_2$ region. 
One of the best studied PDRs is the Orion Bar, which is the paradigm of
a Galactic PDR associated to an HII region. Observations of the H$_2$ rotational
lines in the Orion Bar by \citet{par91} revealed the existence of unexpectedly
large amounts of warm gas (T$\sim$400-700~K). They proposed that clumpiness 
could help to reconcile the observations with chemical models.
Recently, \citet{all05} proposed that the dust FUV attenuation cross sections should be
reduced by a factor of 3 in order to explain the separation between the ionization front
and the H$_2$ emission peak in the Orion Bar. In order to explain the intensities of the
H$_2$ ro-vibrational lines they need to re-adjust the photoelectric heating rate.
A change in the size distribution of the grains and/or the photoelectric heating rate
would produce large variations in the thermal balance of the PDR. 

\acknowledgments

This paper has been partially funded by the Spanish MCyT 
under projects DGES/AYA2000-927, ESP2001-4519-PE, ESP2002-01693, 
AYA2002-01241 and AYA2003-06473. 


Facilities: \facility{30m(IRAM)}.

\clearpage



\begin{figure}
\epsscale{0.8}
\plotone{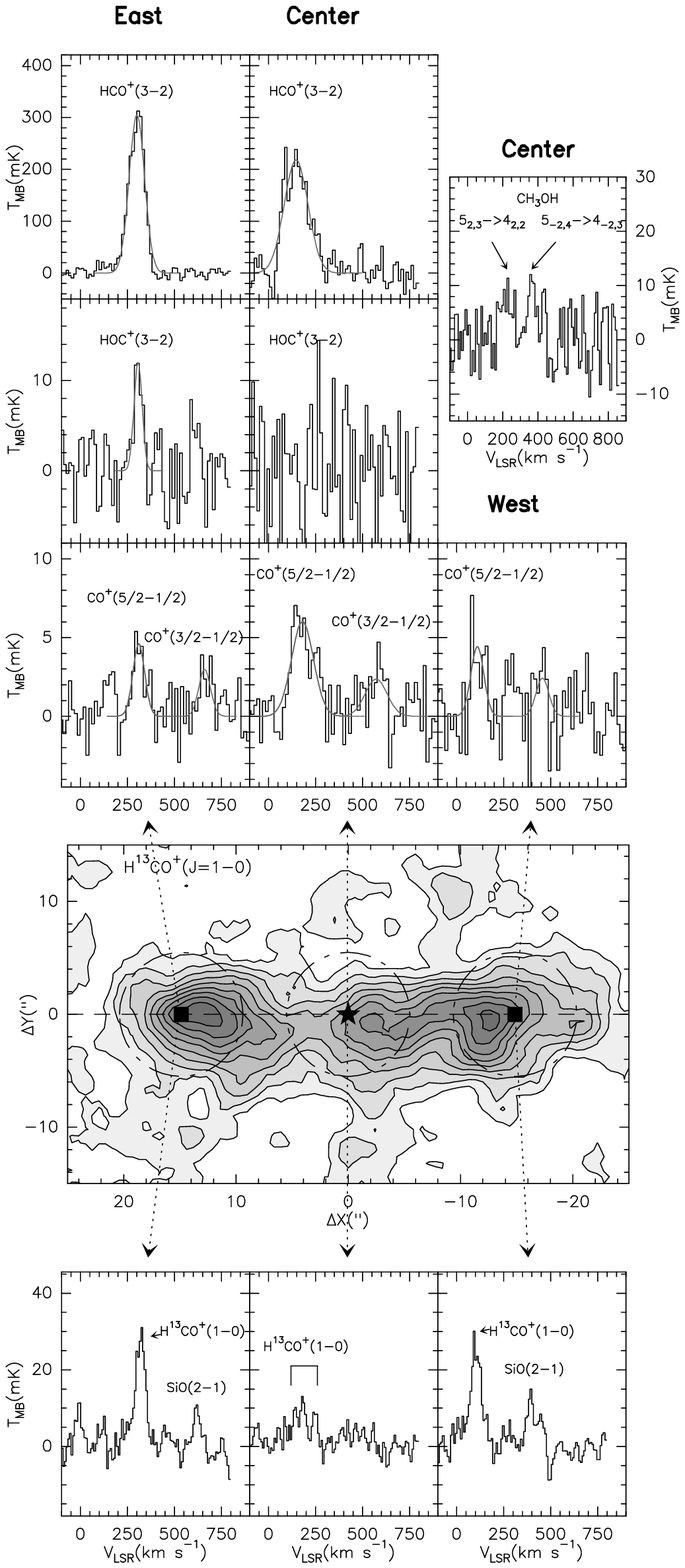}
\caption{Observed spectra towards the positions East  (E), Center and West (W) in M82. The 30m beam at 236 GHz
around the three observed positions has been drawn in the interferometric H$^{13}$CO$^+$ image by \citet{gar02}. 
\label{fig1}}
\end{figure}

\setlength\unitlength{1cm}
\begin{figure*}
\vspace{14cm}
\includegraphics{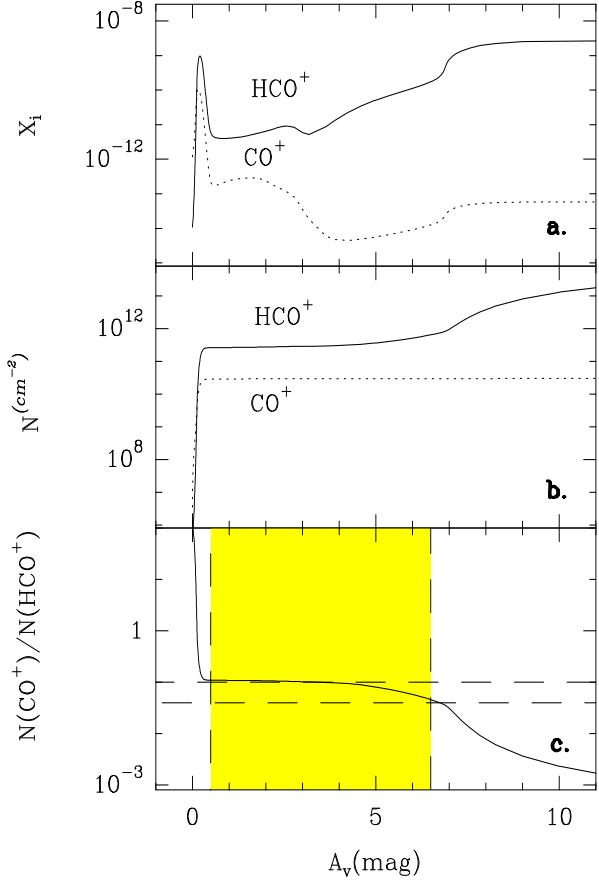}
\caption{Model predictions for the fractional abundances ({\bf a.})and cumulative column densities of CO$^+$ 
and HCO$^+$ ({\bf b.}) derived using the 
Le Bourlot et al.'s code for the physical conditions in the M~82 nucleus (G$_0$=10$^4$ in units of the
Habing field, n=10$^5$~cm$^{-3}$ and $\zeta$=4 10$^{-15}$ s$^{-1}$).  Note that the N(CO$^+$)/N(HCO$^+$)
ratios measured in the M82 nucleus (horizontal dashed lines in {\bf c.} panel) are well explained if the
emission arises in PDRs with total visual extinction between 4.5 and 6.5 mag, in perfect agreement with our
results in Paper I based on the CN/HCN ratio.
\label{fig2}}
\end{figure*}
 
\clearpage
\begin{table}
\begin{center}
\caption{Observational parameters and Gaussian fits results \label{tbl-3}}
\begin{tabular}{lrccccc}
\tableline\tableline
                                 &  \multicolumn{1}{c}{Rest Freq.$^a$} & $\int T_{MB} dv$ &  $v_{lsr}$  &  
 $\Delta v$    &  T$_{MB}$  &  t$_{int}$  \\ 
  Molecule                &   \multicolumn{1}{c}{(MHz)}     &  (K km s$^{-1}$)  &  (km s$^{-1}$)  &   
(km s$^{-1}$)  & (mK)  &   (min)              \\ \tableline
                                 E. (+14$"$,+5$"$)                                                                                                            \\ \tableline
CO$^+$ J=5/2$\rightarrow$3/2       &  236062.55  & 0.36(0.08)  &      311(8)    &   74(19)   & 4.6 & 695 \\
HCO$^+$  J=3$\rightarrow$2          &  267557.00   & 31.33(0.45)  &   301(1)  & 96(2)   & 304.7 & 127 \\
HOC$^+$   J=3$\rightarrow$2          &  268451.00   & 0.65(0.13)    &    307(5)  &  51(12)  & 12.0 & 125 \\
H$^{13}$CO$^+$$^a$  J=1$\rightarrow$0     &  86754.33   &  2.05(0.15)  & 320(2)      & 66(6)    & 29.0 &   \\  \tableline 
                                 Center (0$"$,0$"$)                                                                                                  \\ \tableline
CO$^+$  J=5/2$\rightarrow$3/2        & 236062.55    &  0.88(0.11)  & 181(9)   & 137(19)   &  6.0 &  340  \\
CO$^+$  J=3/2$\rightarrow$1/2       & 235789.64    &  0.34(0.09)  &              & 137$^b$  & 2.4 &  340 \\ 
HCO$^+$   J=3$\rightarrow$2   & 267557.00    & 34.66(1.52)  & 147(4)  & 149(7)    &  218.1  & 12 \\
HOC$^+$     J=3$\rightarrow$2  & 268451.00     & \multicolumn{4}{c}{$<$0.34$^c$ K km s$^{-1}$ } & 190 \\
H$^{13}$CO$^+$$^a$  J=1$\rightarrow$0   & 86754.33  &  1.21(0.13)  & 171(8)      & 140(18)    & 8.1 &  \\   
CH$_3$OH      J$_k$=5$_{2,3}$$\rightarrow$4$_{2,2}$$^d$    &  241904.63   & 
0.92( 0.27) &  222(13) &  85(27)   & 10.0 & 70 \\ \tableline
                                  W. (-14$"$,-5$"$)                                                                            \\ \tableline
CO$^+$    J=5/2$\rightarrow$3/2    & 236062.55      &  0.38(0.10)   &  111(8)  & 81(21)  &  4.4 &  390  \\  
H$^{13}$CO$^+$$^a$    J=1$\rightarrow$0     &  86754.33  & 1.94(0.15)  &  103(3)  &  73(7)   &  24.8 & \\ 
\tableline
\end{tabular}
\end{center}

\noindent
$^a$ Spectra obtained from the interferometric H$^{13}$CO$^+$ image
reported by \citep{gar01} by convolving to a Gaussian
resolution of HPBW=11".

\noindent
$^b$Gaussian fit obtained by fixing the velocity and linewidth. 

\noindent
$^c$3$\times \sigma$ limit with $\Delta$v=50 km~s$^{-1}$

\noindent
$^d$ The line  J$_k$=5$_{-2,4}$$\rightarrow$4$_{-2,3}$ at 241904.15 MHz is overlapped with this.
\end{table}

\begin{table}
{\scriptsize
\begin{center}
\caption{Column densities and relative fractional abundances\label{tbl-3}}
\begin{tabular}{lcccccc}
\tableline\tableline
Molecule                                   & \multicolumn{3}{c}{M82$^1$} & Orion Bar$^2$ & NGC 7023$^3$  & Mon R2$^4$ \\ 
              & E.  & (0,0) & W.  &      IF          &   PDR peak                &    IF    \\ \tableline
N(CN)           &      6.3$\times$10$^{15}$     &   8.8$\times$10$^{15}$    &  1.1$\times$10$^{16}$ &       
2.4$\times$10$^{14}$    &  2.4$\times$10$^{14}$       &       \\
N(CO$^+$)   &  1.5$\times$10$^{13}$  & 3.7$\times$10$^{13}$  & 1.6$\times$10$^{13}$         &
8.0$\times$10$^{11}$   & 1.6$\times$10$^{12}$     &   4.4$\times$10$^{12}$        \\
N(HOC$^+$) & 2.5$\times$10$^{13}$  &  4.3$\times$10$^{13}$ &   2.5$\times$10$^{13}$    & 4.0$\times$10$^{11}$ & 
1.3$\times$10$^{11}$   &    1.4$\times$10$^{12}$         \\
CN/HCN                                     &   6     &    8    &   7      & 3       &   4--8  & ...  \\ 
HCO$^+$/HOC$^+$                    &  44         &  36        &  30   & $<$166    & 50--120  &  460  \\
CO$^+$/HCO$^+$                       & 0.04   & 0.1 & 0.04   &$\sim$0.01 & 0.01--0.11 & 0.005  \\
\tableline
\end{tabular}
\end{center}

\noindent
$^1$ Assuming a size of 6$"$ and data from Paper I and this paper.

\noindent
$^2$ Assuming that the bar fills half of the beam and data from Fuente et al. (1993,1996b).

\noindent
$^3$ Assuming a filament of 6$"$ and data from Fuente et al. (1993,2003).

\noindent
$^4$ Assuming the size of the radio continuum emission at 6 cm, 10$"$, and data from Rizzo et al.(2003).
}
\end{table}
\end{document}